# A Heterogeneity Based Case-Control Analysis of Motorcyclist's Injury Crashes:
# Evidence from Motorcycle Crash Causation Study

Behram Wali
Graduate Research Assistant, Department of Civil & Environmental Engineering
The University of Tennessee
bwali@vols.utk.edu

Asad J. Khattak, Ph.D.
Beaman Distinguished Professor & Transportation Program Coordinator
University of Tennessee, Knoxville, TN 37996, USA
akhattak@utk.edu

Aemal J. Khattak, Ph.D.
Professor and Associate Chair
Department of Civil Engineering
University of Nebraska-Lincoln, NE 68588-6105, USA
khattak@unl.edu

May 10th, 2017

# A Heterogeneity Based Case-Control Analysis of Motorcyclist's Injury Crashes:
# Evidence from Motorcycle Crash Causation Study

Behram Wali[1], Asad J. Khattak[1], Aemal J. Khattak[2]
[1]University of Tennessee, Knoxville
[2]University of Nebraska, Lincoln

**Abstract -** The main objective of this study is to quantify how different "policy-sensitive" factors are associated with risk of motorcycle injury crashes, while controlling for rider-specific, psycho-physiological, and other observed/unobserved factors. The analysis utilizes data from a matched case-control design collected through the FHWA's Motorcycle Crash Causation Study. In particular, 351 cases (motorcyclists involved in injury crashes) are analyzed vis-à-vis similarly-at-risk 702 matched controls (motorcyclists not involved in crashes). Unlike traditional conditional estimation of relative risks, the paper presents heterogeneity based statistical analysis that accounts for the possibility of both within and between matched case-control variations. Overall, the correlations between key risk factors and injury crash propensity exhibit significant observed and unobserved heterogeneity. The results of best-fit random parameters logit model with heterogeneity-in-means show that riders with partial helmet coverage (U.S. DOT compliant helmets with partial coverage, least intrusive covering only the top half of the cranium) have a significantly lower risk of injury crash involvement. Lack of motorcycle rider conspicuity captured by dark (red) upper body clothing is associated with significantly higher injury crash risk (odds ratio 3.87, 95% CI: 1.63, 9.61). Importantly, a rider's motorcycle-oriented lower clothing (e.g., cannot easily get stuck in the machinery) significantly lowers the odds of injury crash involvement. Regarding the effectiveness of training, formal motorcycle driving training in recent years was associated with lower injury crash propensity. Finally, riders with less sleep prior to crash/interview exhibited 1.97 times higher odds of crash involvement compared to riders who had more than 5 hours of sleep. Methodologically, the conclusion is that the correlations of several rider, exposure, apparel, and riding history related factors with crash risk are not homogeneous and in fact vary in magnitude as well as direction. The study results indicate the need to develop appropriate countermeasures, such as refresher motorcycle training courses, prevention of sleep-deprived/fatigued riding, and riding under the influence of alcohol and drugs.

## 1. INTRODUCTION & BACKGROUND

Recent statistics suggest that the annual number of motorcycle fatalities has increased by 48 percent since 2002, i.e., from 3,365 fatalities in 2002 to 4,976 fatalities in 2015 (NHTSA, 2016). This increase discernibly contrasts with a 32-percent decrease in fatalities to occupants of passenger cars and light trucks (NHTSA, 2016). Alarmingly, after accounting for per vehicle mile traveled, motorcyclists are fatally injured 29 times more frequently than their passenger vehicle counterparts (NHTSA, 2016). Owing to the expanding concern, the U.S. Congress recently passed a legislation to initiate the most comprehensive research effort targeted at identifying the causes of motorcycle crashes (NHTSA, 2017). As motorcycle safety remains a significant concern, different prevailing themes exist in the research arena. Extensive literature is focused on lowering the occurrence and unsafe outcomes of motorcycle crashes (Preusser et al., 1995, Sosin et al., 1990, Haque et al., 2010, Quddus et al., 2002, Savolainen and Mannering, 2007, Chin and Quddus, 2003). The key focus of the crash frequency and/or injury severity literature is to investigate how a confluence of factors may contribute to the occurrence and/or outcomes of motorcycle related crashes.

To link motorcyclist's injury outcomes with explanatory factors, discrete outcome models are typically used such as multinomial logit (Shankar and Mannering, 1996), nested logit (Savolainen and Mannering, 2007), mixed logit (Shaheed et al., 2013), and ordered probit models (Quddus et al., 2002, Blackman and Haworth, 2013). Collectively, a variety of crash contributory factors such as driver-related factors (Schneider et al., 2012), roadway geometrics (Quddus et al., 2002, Haque et al., 2010), motorcycle characteristics (Savolainen and Mannering, 2007), and environmental factors (Rifaat et al., 2012) are identified to be associated with motorcycle crash outcomes. While such an analysis provides valuable insights into understanding the motorcyclists' injury outcomes, it does not shed light on the risk-taking behaviors of motorcyclists and how it relates to crash risk.

Another area of continuing research is the motorcyclist's crash risk and its associated factors. The behavioral literature focusing on motorcyclists attitudes toward road safety suggest that motorcyclists are less risk averse than other motorized road users (Fuller et al., 2008, Broughton et al., 2009). As motorcyclists are one of the most crash-prone groups, an understanding of how different behavioral, situational, and rider-specific factors relate to crash risk has also been sought. Several high-risk riding behaviors such as speeding, riding under influence, non-usage of helmets, unlicensed riding, and inexperience are correlated with higher crash rates (Lin et al., 2003, Schneider et al., 2012, Elliott et al., 2007, Moskal et al., 2012, Rowden et al., 2016). Keall and Newstead found that crash rates for motorcycles were four times that of passenger cars (Keall and Newstead, 2012). Note that the crash rate per vehicle for motorcycles was found to be only 30% higher than that for small vehicles, however, when adjusted for distances travelled, the crash rate for motorcycles became almost four times that of small vehicles (Keall and Newstead, 2012). Compared to older riders, riders aged 20-29 years had significantly higher crash rates (Keall and Newstead, 2012). Regarding high-risk behaviors, the odds of crash involvement were three times higher for riders who reported "frequent stunt" behaviors (Stephens et al., 2017). Likewise, a positive relationship was reported between self-reported traffic errors, control errors, and motorcycle crashes or crash liability (Cheng et al., 2015, Elliott et al., 2007). For the most part, the analysis of rider-specific behavioral factors, socioeconomic factors, demographics, and motorcycle driving experience related factors correlated with motorcyclist crash risks mainly builds upon data from the traditional police crash reports or questionnaire surveys (Schneider et al., 2012, Stephens et al., 2017, Elliott et al., 2007). While useful, such an analysis usually does not reflect the exposure of the population under study (i.e., motorcyclists) to the outcome of interest (i.e., motorcyclist crash). Also, insights regarding the interrelationships between explanatory factors and actual crash propensity cannot be easily obtained.

Crash propensity is defined as the likelihood of rider's involvement in injury crash events, compared to non-crash events. In particular, compared to non-crash events, how different trip related factors, physical, psychological, and exposure-related factors relate to the likelihood of motorcycle rider's involvement in injury crash events? An analysis of differences in the behavioral, situational, and rider-specific factors of motorcyclists involved in injury crash and non-crash events can ultimately facilitate formulation of actionable crash prevention countermeasures. A case-control data structure is needed for quantification of risk of associated factors on crash propensity. That is, detailed data are needed for riders involved in crash and non-crash events. To the authors knowledge, no comprehensive, large-scale study of motorcycle crashes had been conducted in the U.S. since the commonly called Hurt Study (Hurt et al., 1981), which included an on-scene investigation of motorcycle crashes in Los Angeles between 1976 and 1980 (Hurt et al., 1981). Thus, due to non-availability of suitable data, meager evidence exists in published literature regarding relative risks of key behavioral and crash specific factors, and how such key-risk factors relate to motorcycle crash propensity (i.e., involvement of a motorcycle rider in an injury crash vs. a non-crash event). Such an analysis is central to the formulation of actionable countermeasures for preventing motorcycle crashes.

With these forethoughts in mind, the main objective of this study is to quantify how different behavioral, psycho-physiological, and exposure-related factors relate to the likelihood of rider's involvement in injury crash events. To achieve this, a matched case-control study design is adopted to calculate relative risks of several exogenous factors. A unique aspect of this study is the incorporation of matched controls (motorcyclists not involved in crashes) which provide a basis for comparisons of rider and vehicle characteristics. The controls are matched with the case events (motorcyclists involved in injury crashes) for time of day, weather, road type, urban/rural, and other factors, and thus can be regarded as similarly-at-risk controls. From a methodological standpoint, unlike the traditional conditional estimation of relative risks, comprehensive heterogeneity based statistical analysis is conducted that accounts for the possibility of both within- and between matched case-control variations. In addition, heterogeneity in the relative risks, both due to observed and unobserved factors, is addressed. Compared to the commonly-used random parameters models that typically assume the same mean for each random parameter, the study accounts for possible heterogeneity in the means of the random parameters which vary as a function of several observed factors. To the best of authors' knowledge, the use of such a method has not been used/reported in a retrospective matched case-control design context.

## 2. METHODOLOGY

### 2.1. Development of Case-Control Strategy

The notion of *crash propensity* is the likelihood of the motorcyclist's involvement in an injury crash event. From here onwards, the term "crash propensity" is used to refer to the risk of rider's involvement in an injury crash. *Risk factors* refer to the explanatory variables associated with an increased likelihood of motorcycle crash. Crash frequency data are typically analyzed to understand risk factors associated with motorcycle crashes (Chin and Quddus, 2003, Schneider et al., 2012, Blackman and Haworth, 2013). While useful, such an analysis usually does not reflect the exposure of the population under study (i.e., motorcyclists) to the outcome of interest (i.e., motorcyclist crash). To circumvent this, a retrospective matched case-control design is adopted in this research to better understand the association of risk factors with the motorcyclist crash propensity (Figure 1). The two units that assemble a case-control design are the *cases* and *controls*. Cases are riders involved in injury crashes during a specific time-period, whereas, controls are riders that are not involved in crashes during the same time while exhibiting similar exposure as their case counterparts. The controls provide a basis for comparison of motorcycle, environment, and rider characteristics. To better understand crash propensity while accounting for overall exposure of the population (cases and controls in this case), each case is matched with two controls by time of day, day of week, weather, road type, urban/rural, location, and travel direction (Figure 1). Importantly, the controls in this study can be regarded as "similarly-at-risk" controls. Once controls are generated, they can be coupled with cases to formulate a binary response outcome which can then be modeled using appropriate statistical methods. With such a 1:2 matched case-control study design, collecting data on exposure and other specific factors eventually allows quantification of relative risks that may be useful for countermeasure development. At a basic level, the comparison of the frequency of key factors in cases and controls can spot differential involvements.

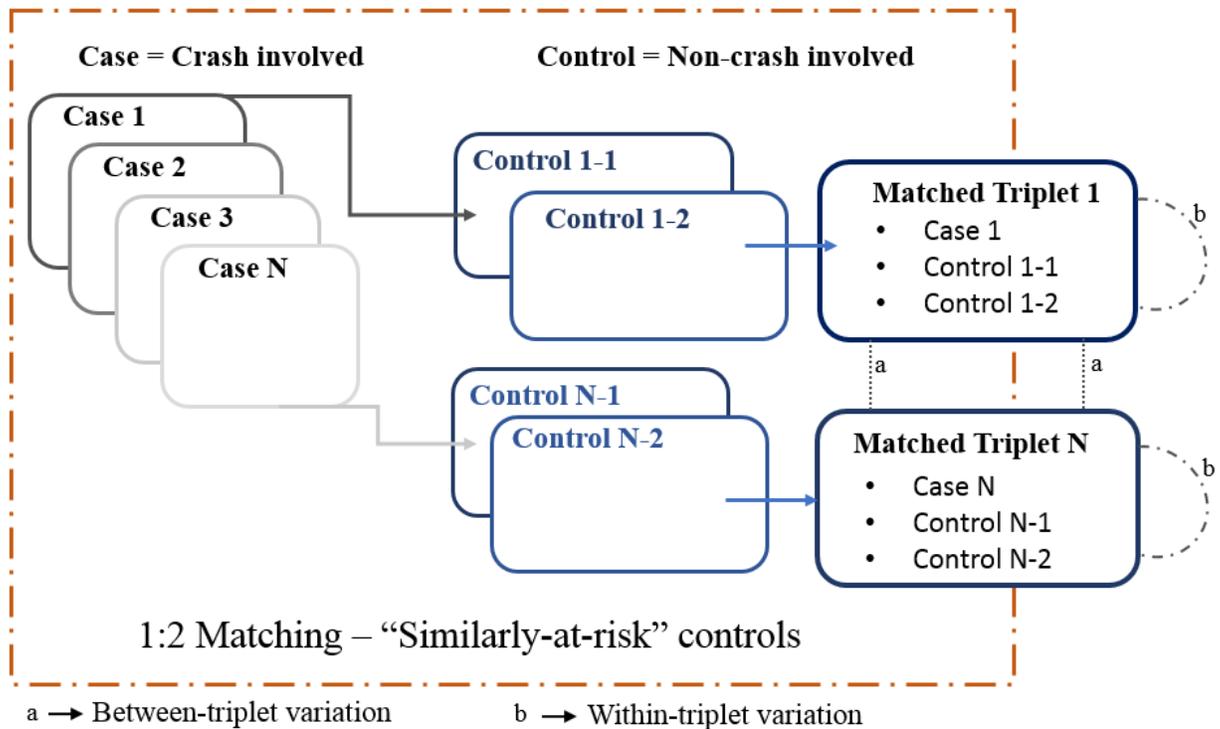

**FIGURE 1: A Retrospective Matched Case-Control Approach**

## 2.2. Data Source

The methodological framework shown in Figure 1 is inspired by epidemiological and ecological research (Barzilai et al., 2003, Atzmon et al., 2004). The proposed approach utilizes their methodological strengths in systematic analysis of case-control approaches by pairing presence and absence of certain disease (injury crashes in this case). The pairing of cases (e.g., presence of disease) and controls (absence of disease) leads to development of strata, which is a matched triplet in the case of this research (Figure 1). The retrospective matched case-control approach shown in Figure 1 builds upon the Federal Highway Administration (FHWA) Motorcycle Crash Causation Study (MCCS) (FHWA, 2017a). Importantly, MCCS is the most comprehensive data collection effort in the United States in more than 30 years legislated by the U.S. Congress and sponsored by the U.S. Department of Transportation (FHWA, 2017a). The dataset includes data from comprehensive on-scene investigations of 351 motorcycle injury crashes in Orange County, California, and 702 control rider interviews[1,2] (FHWA, 2017a).

---

[1] Note that the FHWA's MCCS data collection is complete, the FHWA is reviewing a final report and detailed study documentation. According to FHWA website (https://www.fhwa.dot.gov/research/tfhrc/projects/safety/motorcycles/mccs/), the final report and volumes 1 and 2 will be available in 2018. While the data and draft versions of the documentation are freely available from the FHWA upon request (see (FHWA, 2017a)), documentation of the specific volumes of data used in this study (see footnote 2) can be requested from the authors.

[2] MCCS is an extensive data collection effort where the draft final report consists of 14 volumes related to different data elements. Given the focus of the present study, we mainly used Volume 2 (crash form data) and Volume 5 (motorcycle crash and control rider data) of the MCCS data for analysis. The crash form data provides information about 15 data elements attributes for each of the 351 crashes involved in the study. The motorcycle crash and control rider data provides information about the crash driver's responses to 105 questions and the control rider's responses to 88 questions. The same questions are given to both the crash and the control parties, but some are unique to the crash-involved motorcycle riders.

## 2.3. Crash & Control Case Data Collection Protocol & Comparability

For each of the injury crash, comprehensive data are available including pre-crash, crash, and post-crash characteristics of the riders and crash sites. Environmental details, traffic controls, and roadway features that may have contributed to crash causation are obtained to get a fuller picture of the crash event (FHWA, 2017b). Also, the MCCS team collected data on injury details and human factors such as medical conditions, fatigue levels, demographics, and trip purposes through direct observations, medical records available to the investigation team, and interviews. The crash and control data collection protocol is detailed and sophisticated, and it is impractical to cover all of the specific details of the protocol; for parsimony we provide an overview. For details see (FHWA, 2017b) and footnotes 1 and 2. Due to the well-known intrinsic limitations of police-reported crash data, injury and other information provided in police-reported crash data can be subjective and vulnerable to bias (Wali et al., 2018e, Mannering and Bhat, 2014). To address these issues to the largest extent possible, the MCCS uses unique and rigorous protocols to collect injury data for crash involved riders. It is important to note that injury data reported in the MCCS is not solely obtained from police-reports, but rather comprehensive descriptions of all injuries (including minor) obtained by trained investigators through crash-involved motorcyclists' occupant interviews. Because access to medical information is carefully controlled through the U.S. Federal Health Insurance Portability and Accountability Act (HIPAA), investigators executed signed patient release forms in order to obtain copies of patient injury records, emergency room reports, patient discharge summaries, and medical records from private physicians (if applicable) (FHWA, 2017a). Importantly, as autopsies are public records in California (which is the MCCS study was conducted), the medical examiner also provided autopsy reports if applicable. Investigators then examined and encoded these exhaustive sets of records.

As mentioned earlier, to help compare the characteristics and the associated risk factors of crash-involved riders and non-crash involved riders, the MCCS gathered control data through detailed interviews with motorcycle riders who were similarly at risk to those involved in crashes. The MCCS study followed the principles of collecting detailed data by studying two non-crash-involved "control" motorcyclists for each focal crash observed during the study period (Figure 1). Thus, the control cases provide a unique basis for comparing operator/rider and vehicle characteristics. The validity of this study relies on the comparability of the control and crash data. That is, can the control riders be considered as "similarly-at-risk" with their crash counterparts? The control data collected by the MCCS team are comparable to the crash data and can be regarded as "baselines". The control data (baselines) are comparable because for each crash the two control subjects were matched with corresponding crash subjects based on 1) location, 2) road type, 3) urban/rural, 4) travel direction, 5) day of week, 6) time of day, and other factors (FHWA, 2017a). The matched time period covers approximately one hour before and after the reported time of the motorcycle crash. In addition, as explained in footnote # 2, the fact that the majority of variables are common to both crash and control groups further enhances the comparability of control data. Collection of exposure data on control riders matched this way (i.e., similarly-at-risk controls) makes it comparable with the crash data. Because several of the factors listed above are common between the crash and control riders, the data allows for the calculation of relative risks of factors that can increase the chance of a crash (compared to baselines), and thus can be used for developing countermeasures

As mentioned earlier, to help compare the characteristics and the associated risk factors of crash-involved riders and non-crash involved riders, control data are gathered through detailed interviews with motorcycle riders who were similarly at risk to those involved in crashes. The MCCS study followed the principles of collecting detailed data on two non-crash-involved "control" motorcyclists for each focal crash observed during the study period (Figure 1). Thus, the control cases provide a unique basis for comparing operator/rider and vehicle characteristics.

The MCCS used two methods, that are approved by the Oklahoma State University's IRB, for acquiring the matched controls (FHWA, 2017b):
1. Voluntary traffic stops at or near the crash scene (same time of day, day of week, and direction of travel); and,
2. Recruiting motorcyclists who may be at nearby gas stations.

The MCCS established control data collection sites at locations that were at or near the related crash site. The control data collection team notified concerned local police jurisdiction of the time and locations of the data collection sites. At each site, the team displayed proper signage inviting volunteer motorcycle riders and passengers to participate in the MCCS. The signage also indicated that a $40 gas card would be given to the volunteer riders (FHWA, 2017b). If two controls for each focal injury crash were unavailable at a particular time for a particular crash, then the procedure was repeated at the same location during a second-time period. Finally, interviews with the volunteer motorcycle riders and inspections of their vehicles were conducted for all 702 controls. A critical element of the data collection process is that data related to demographics, riding experience, formal motorcycle driving training, riders' clothing and safety equipment, type and coverage of helmet usage, traffic convictions, and other factors is collected both for crash-involved riders (cases) and non-crash involved riders (controls) (Table 1). In addition, information about trip purpose, speed before the crash/interview, possible impairments, and, for control riders who agreed, data on volunteer breath tests for Blood Alcohol Concentration (BAC) are also available (Table 1). In their attempts to verify and document all of the coded data elements, the trained investigators also inspected and photographed the vehicles of control riders, as was done with the crash-involved motorcycles. The MCCS is conducted in strict adherence to Institutional Review Board (IRB) standards. For details see (FHWA, 2017a).

Overall, the MCCS study was rigorously designed and carefully executed. As an example, a Pilot Study that preceded the MCCS study in order to test and validate the study's design developed the necessary training materials and relationships with local law enforcement agencies, data collection and coding forms, and developed and assessed research protocol designed for the main study (MCCS). The Pilot Study conducted in 2008 completed a total of 23 crash investigations conducted in Orange County, California (NHTSA, 2010). As documented in detail in the National Highway Traffic Safety Administration's (NHTSA) Pilot Study report, the conclusion of the Pilot Study made a few minor edits to some of the data forms and the related sections in the coding manual (NHTSA, 2010).

## 2.4. Empirical Approach

The analysis of matched case-control requires application of statistical methods that can account for the matching structure in the data. For instance, matching of the two controls with each focal injury crash (case) by common matching characteristics results in a dependency amongst the three units within each triplet (Figure 1). As such, given the binary nature of the response outcome (crash involved rider vs. non-crash involved rider), a widely used approach in the literature for analysis of case-control data is the conditional logistic regression, also known as fixed-effects logit model (Hartman et al., 2004, Meuleners et al., 2015, Rothman et al., 2017). Compared to the unconditional logistic model, the fixed-effects logit framework accounts for the within-triplet variation and dependence (indicated by "b" in Figure 1). However, the traditional fixed-effects logit framework does not account for an important methodological concern, i.e., the possibility of between-observation or between-triplet heterogeneity due to unobserved factors systematically varying across the individual observations or the matched triplets (indicated by "a" in Figure 1) (Wali et al., 2018b, Li et al., 2017). The very likely possibility of heterogeneity due to systematic variation in unobserved factors *(i.e., unobserved heterogeneity)* is an outgrowth of a confluence of interdependent factors related to rider characteristics, roadway factors, exposure-related factors, environmental and weather-related factors, all of which are associated with crash occurrence. Thus, for drawing a fuller picture of factors associated with injury crash propensity, the unobserved important factors should be accounted for in the analysis (Zhao and Khattak, 2017, Wali et al., 2017, Liu et al., 2017, Shaheed et al., 2013, Khattak and Wali, 2017). Nonetheless, data on all the factors known to affect motorcyclist crashes are not available (Mannering et al., 2016, Wali et al., 2018a, Boakye et al.). Having said this, we account for unobserved heterogeneity in the modeling specification by estimating fixed and random parameter logit models.

A second important concern relates to the issue of *observed heterogeneity*, i.e., variations in the associations between exogenous variables and crash propensity due to systematic variation in the observed factors. While models accounting for unobserved heterogeneity have become a methodological standard in the safety literature (not in the case-control literature though) (Wali et al., 2018d, Wali et al., 2018c), a typical approach is to fix the mean parameter estimate for random-parameters across different units (individual observations or matched-triplets) (Mannering and Bhat, 2014, Li et al., 2017). Within the random-parameter framework, accounting for observed heterogeneity by using appropriate methods can further help in extracting deeper insights from the data (Wali et al., 2018d). We briefly present the model specifications of fixed and random parameter logit models that account for both observed and unobserved heterogeneity in the context of motorcycle crash propensity analysis.

### *2.4.1. Crash Propensity Models at Individual Observation Level:*

Following the data structure in Figure 1, comprehensive heterogeneity based statistical analysis is conducted both at individual observation level and at matched-triplet level (discussed later). The fixed and random parameter crash propensity models described here operate at the individual observation level, i.e., considering cases and controls to be independent. Let $Y_i^*$ be the latent crash propensity of rider $i$ that is unobserved, and can be specified as (Boakye et al., 2018, Wali et al., 2018d, Tay, 2016):

$$Y_i^* = X_i\beta + u_i \qquad (1)$$

Where: $X_i$ is a vector of rider-specific explanatory variables, $\beta$ represents a vector of parameter estimates, and $u_i$ is a random disturbance term (Boakye et al., 2018, Tay, 2016). In this case, the observed response outcome is dichotomous variable, such that, $Y_i = 1$ represents injury crash involved rider, and $Y_i = 0$ represents non-crash involved rider, and:

$$Y_i = \begin{cases} 1 \text{ if } Y_i^* > \mathbf{0} \\ 0 \text{ otherwise} \end{cases} \qquad (2)$$

The probability of rider $i$ getting involved into a crash then becomes:

$$\begin{aligned} P_i &= prob[X_i\beta + u_i > 0] \\ P_i &= prob[u_i > (X_i\beta)] \\ P_i &= 1 - F(-X_i\beta) \end{aligned} \qquad (3)$$

Where F is the cumulative density function of $u$. To account for the unobserved heterogeneity in the standard logit framework, random parameters can be introduced as (Wali et al., 2018a, Fountas and Anastasopoulos, 2017):

$$\beta_i = \beta + \mathbf{Y}\zeta_i \qquad (4)$$

Where: $\beta$ is the mean of random parameter vectors (either intercept or exogenous variables), $\mathbf{Y}$ is the diagonal matrix with standard deviations for random parameters, and $\zeta_i$ is a normally distributed term with mean zero and variance one (Wali et al., 2018a, Li et al., 2017, Behnood and Mannering, 2017, Mannering et al., 2016). The standard fixed parameter model shown in Equation 3 can be estimated by maximum likelihood estimation method. Whereas, the estimation of Equation 4 proceeds with Maximum Simulated Likelihood procedures where scrambled Halton draws (compared to random draws) are used in the simulation process. In this study, 1000 Halton draws are used for parameter estimation, nonetheless, 200 Halton draws are reported to produce accurate parameter estimates (Bhat, 2003). Regarding function form of the parameter density functions, we tested normal, lognormal, triangular, uniform, and Weibull distributions. Further details about random parameter modeling can be found in (Boakye et al., 2018, Wali et al., 2018b, Khan and Khattak, 2018, Zhao and Khattak, 2015, Bhat, 2003).

While the mathematical formulation in Eq. 4 accounts for unobserved heterogeneity by estimating different set of event-specific parameter estimates $\beta_i$, nonetheless, the mean parameter estimate ($\beta$) is still fixed across all the observations. Unlike commonly-used random parameters models shown above that typically assume the same mean for each random parameter, we also allow the means of random parameters to vary across observations as a function of observed explanatory factors. Thus, Eq. 4 becomes (Wali et al., 2018d, Behnood and Mannering, 2017):

$$\beta_i = \beta + \xi Z_i + Y\zeta_i \tag{5}$$

Where: $\beta$ is the mean parameter estimate across all crashes $i$, $Z_i$ is a vector of explanatory factors from observation $i$ which influence the mean of $\beta_i$, $\xi$ is the parameter vector associated with $Z_i$, and $Y\zeta_i$ are as defined earlier that accounts for unobserved heterogeneity across different crashes. In addition to accounting for unobserved heterogeneity, the formulation in Eq. 5 now also accounts for observed heterogeneity by allowing the means of random parameters to vary as a function of specific observed factors (Wali et al., 2018d).

### 2.4.2. Crash Propensity Models at Matched-Triplet Level:

The different fixed and heterogeneity (both observed and unobserved) based model specifications shown in Equation 1 through 5 operate at individual observation level. However, there exists a matching structure in the data in the sense that the two controls are matched with each focal crash (case) by common matching characteristics. This results in a dependency among the three units within each matched triplet (Figure 1). As such, the heterogeneity framework discussed above should be enhanced to account for the dependency structure within each triplet. For the standard random parameter model at matched-triplet level, Equation 4 now becomes:

$$\varrho_j = \varrho + M\Gamma_j \tag{6}$$

Where: $\varrho$ is the mean of random parameter vectors at the matched-triplet level $j$, $M$ is the diagonal matrix with standard deviations for random parameters, and $\Gamma_j$ is a normally distributed term with mean zero and variance one. To further account for observed heterogeneity in the matched-triplet level model specification, the mean random parameters $\varrho$ can be allowed to vary across each matched-triplet as a function of explanatory factors:

$$\varrho_j = \varrho + \delta G_i + \eta\varsigma_j \tag{7}$$

For estimation of matched-triplet level heterogeneity models, 1000 Halton draws are used for parameter estimation, whereas the distributions tested for random parameters are normal, lognormal, triangular, uniform, and Weibull distributions. For comparing the different models using individual and matched-triplet level data, goodness-of-fit measures such as likelihood ratio test, Akaike Information Criteria (AIC), Finite sample AIC, and McFadden Pseudo $R^2$ are used (Washington et al., 2010).

## 3. RESULTS
### 3.1. Descriptive Statistics

At a basic level, the comparison of the frequency of key factors in cases (injury crash-involved riders) and controls (non-crash involved riders) can spot differential involvements. As such, Table 1 presents the descriptive statistics of key variables for the crash and control group. In total, 351 cases and 702 controls are analyzed (Table 1). Also, the results of two-sample t-tests (conducted at 95% confidence level) for the crash and control groups are presented to spot statistically significant differences in the means of particular risk factors across the two groups (Table 1).

Regarding the results of the two-sample t-tests, a "pass" in Table 1 indicates that the null hypothesis cannot be rejected at 95% confidence level, i.e., the two means are statistically not different, whereas a "fail" indicates that the null hypothesis can be rejected at 95% confidence level, i.e., the two means are statistically different. Compared with non-crash involved riders (control), the trips of injury crash involved riders were statistically significantly less frequently originating at home, and more frequently originating at work or friend/relative place (Table 1). Compared to the control group, using a road rarely (road used once per week and road used once per month) was less frequent for injury crash-involved riders. Regarding motorcycle helmet coverage, compared to the control group, riders with helmet coverage type 1 (US DOT compliant helmets with partial coverage, least intrusive covering only the top half of the cranium) were less frequent in the injury crash group. Likewise, riders with acceptable helmet fit were also less frequent in the injury crash group. The finding that riders with partial helmet coverage are less frequent in crash group is intuitive; as full coverage helmets may interfere with the rider's hearing and vision capabilities, and thus may increase probability of crash. However, note that helmets are pivotal in reducing head injuries, given a crash (Brandt et al., 2002). Regarding overall helmet use, 100% of the control group riders were wearing helmets at the time of the interview. The helmet use among crash group riders was also high, i.e., riders in only 4 crashes were not wearing helmet at the time of crash.

Related to physical and psychological factors, riders with no physical and psychological impairments were statistically significantly less frequently involved in crashes (Table 1). Regarding speed before motorcycle crash or interview, the average rider speed is greater for the control-group riders.

Table 1 also shows the differences related to experience and exposure-related factors. Compared to control (non-crash involved) group, riders with greater years of riding experience, and riders with larger number of miles driven appear less frequently in the crash group. Likewise, riders who completed state recognized entry-level riding course, and riders who received formal training after year 2000 appear more frequently in the control group, i.e., no crash involvement (Table 1). Importantly, compared to control group, riders with four or more traffic convictions in past 5 years appear statistically significantly more frequently in the crash group (mean of 0.52 in crash group vs. mean of 0.04 in control group) (Table 1).

**TABLE 1: Descriptive Statistics of Key Variables**

| Variables | Crash Group (N = 351) | | | Non-Crash Group (N = 702) | | | Mean Comparison Test --- Ho: $\mu_2 - \mu_1 = 0$ |
|---|---|---|---|---|---|---|---|
| | $\mu_1$ | SD | Min/Max | $\mu_2$ | SD | Min/Max | |
| **Trip Origin/Destination** | | | | | | | |
| Origin: Home[a] | 0.30 | 0.46 | 0/1 | 0.89 | 0.32 | 0/1 | Fail |
| Origin: Work[a] | 0.11 | 0.31 | 0/1 | 0.06 | 0.25 | 0/1 | Fail |
| Origin: Friend/relative place[a] | 0.07 | 0.26 | 0/1 | 0.01 | 0.12 | 0/1 | Fail |
| Destination: Friend/relative place[a] | 0.08 | 0.27 | 0/1 | 0.04 | 0.20 | 0/1 | Fail |
| *Frequency of road use* | | | | | | | |
| First-time use[a] | 0.04 | 0.20 | 0/1 | 0.04 | 0.19 | 0/1 | Pass |
| Daily road use[a] | 0.33 | 0.47 | 0/1 | 0.37 | 0.48 | 0/1 | Pass |
| Road used once per week[a] | 0.10 | 0.30 | 0/1 | 0.34 | 0.47 | 0/1 | Fail |
| Road used once per month[a] | 0.04 | 0.20 | 0/1 | 0.20 | 0.40 | 0/1 | Fail |
| Road used once per quarter[a] | 0.01 | 0.12 | 0/1 | 0.03 | 0.18 | 0/1 | Pass |
| *Type of helmet coverage* | | | | | | | |
| Helmet coverage type 1 (Partial coverage)[a] | 0.12 | 0.32 | 0/1 | 0.32 | 0.47 | 0/1 | Fail |
| Helmet coverage type 2 (Full coverage)[a] | 0.03 | 0.18 | 0/1 | 0.04 | 0.19 | 0/1 | Pass |
| Helmet coverage type 3 (full facial, retractable chin bar)[a] | 0.04 | 0.20 | 0/1 | 0.05 | 0.22 | 0/1 | Pass |
| Helmet coverage type 4 (full facial, integral chin bar and face shield)[a] | 0.47 | 0.50 | 0/1 | 0.50 | 0.50 | 0/1 | Pass |
| Helmet fit (1 if acceptable fit, 0 otherwise)[a] | 0.50 | 0.50 | 0/1 | 0.94 | 0.24 | 0/1 | Fail |
| *Physical/psychological factors* | | | | | | | |
| No physical impairment[a] | 0.35 | 0.48 | 0/1 | 0.77 | 0.42 | 0/1 | Fail |
| No psychological impairment[a] | 0.44 | 0.50 | 0/1 | 0.81 | 0.39 | 0/1 | Fail |
| Hours of sleep prior to event | 7.67 | 1.24 | 2/12 | 8.12 | 1.75 | 1/16 | Fail |
| *Drug use* | | | | | | | |
| Alcohol use[a] | 0.13 | 0.34 | 0/1 | 0.20 | 0.40 | 0/1 | Fail |
| Drugs use[a] | 0.09 | 0.28 | 0/1 | 0.16 | 0.37 | 0/1 | Fail |
| Alcohol and multiple drugs[a] | 0.03 | 0.17 | 0/1 | 0.03 | 0.18 | 0/1 | Pass |
| Depressant[a] | 0.03 | 0.16 | 0/1 | 0.03 | 0.17 | 0/1 | Pass |
| **Exposure-related factors** | | | | | | | |
| Motorcycle riding experience in years | 11.52 | 13.63 | 0/46 | 20.48 | 17.07 | 0/69 | Fail |
| Average number of days per year of riding the motorcycle | 225.47 | 116.32 | 3/365 | 199.42 | 110.47 | 4/365 | Fail |
| One-way trip mileage | 19.80 | 23.84 | 1/150 | 23.92 | 42.90 | 1/600 | Pass |
| Total miles driven prior to event | 10.35 | 16.63 | 1/96 | 19.05 | 33.83 | 1/600 | Fail |
| Gaps exist between motorcycle driving[a] | 0.19 | 0.39 | 0/1 | 0.42 | 0.49 | 0/1 | Fail |
| Duration of gap in days | 3.08 | 5.92 | 0/30 | 3.95 | 7.35 | 0/55 | Pass |
| Percentage of driving motorcycles vs diving other vehicles | 55.45 | 33.55 | 0/100 | 47.64 | 31.94 | 0/100 | Fail |

Notes: N is sample size; $\mu_1$ is mean of crash group; $\mu_2$ is mean of control group; SD is standard deviation; Ho indicates the null hypothesis; (a) indicates indicator variables (1/0); Pass indicates that the null hypothesis cannot be rejected at 95% confidence level; Fail indicates that the null hypothesis can be rejected at 95% confidence level.

**TABLE 1: Descriptive Statistics of Key Variables** *(Continued)*

| Variables | Crash Group (N = 351) | | | Non-Crash Group (N = 702) | | | Mean Comparison Test --- Ho: $\mu_2 - \mu_1 = 0$ |
|---|---|---|---|---|---|---|---|
| | $\mu_1$ | SD | Min/Max | $\mu_2$ | SD | Min/Max | |
| *Type & Year of training* | | | | | | | |
| Training type 1 (state recognized, entry-level motorcycle course)[a] | 0.27 | 0.44 | 0/1 | 0.45 | 0.50 | 0/1 | Fail |
| Training type 2 (self-taught)[a] | 0.03 | 0.17 | 0/1 | 0.18 | 0.39 | 0/1 | Fail |
| Training type 3 (taught by family/friend)[a] | 0.03 | 0.18 | 0/1 | 0.07 | 0.25 | 0/1 | Fail |
| Training between 2001-2010[a] | 0.09 | 0.29 | 0/1 | 0.29 | 0.45 | 0/1 | Fail |
| Training between 2011- 2015[a] | 0.07 | 0.25 | 0/1 | 0.20 | 0.40 | 0/1 | Fail |
| *Number of traffic convictions in last 5 years* | | | | | | | |
| One traffic conviction[a] | 0.16 | 0.37 | 0/1 | 0.23 | 0.42 | 0/1 | Fail |
| Two traffic convictions[a] | 0.11 | 0.31 | 0/1 | 0.10 | 0.30 | 0/1 | Pass |
| Three traffic convictions[a] | 0.03 | 0.18 | 0/1 | 0.04 | 0.19 | 0/1 | Pass |
| Four or more traffic convictions[a] | 0.52 | 0.50 | 0/1 | 0.04 | 0.21 | 0/1 | Fail |
| *Rider's apparel* | | | | | | | |
| Upper clothing motorcycle oriented[a] | 0.32 | 0.47 | 0/1 | 0.44 | 0.50 | 0/1 | Fail |
| Lower clothing motorcycle oriented[a] | 0.05 | 0.21 | 0/1 | 0.25 | 0.44 | 0/1 | Fail |
| Shoes motorcycle oriented[a] | 0.16 | 0.37 | 0/1 | 0.31 | 0.46 | 0/1 | Fail |
| Gloves motorcycle oriented[a] | 0.01 | 0.12 | 0/1 | 0.64 | 0.48 | 0/1 | Fail |
| Upper clothing retroreflective[a] | 0.13 | 0.34 | 0/1 | 0.20 | 0.40 | 0/1 | Fail |
| *Clothing color – Torse* | | | | | | | |
| No dominating color, multi-colored[a] | 0.01 | 0.12 | 0/1 | 0.04 | 0.20 | 0/1 | Fail |
| White[a] | 0.04 | 0.20 | 0/1 | 0.07 | 0.26 | 0/1 | Pass |
| Black[a] | 0.34 | 0.47 | 0/1 | 0.52 | 0.50 | 0/1 | Fail |
| Red[a] | 0.04 | 0.20 | 0/1 | 0.03 | 0.18 | 0/1 | Pass |
| *Type of operator's license* | | | | | | | |
| Leaner's permit only[a] | 0.03 | 0.18 | 0/1 | 0.02 | 0.13 | 0/1 | Pass |
| Motorcycle license[a] | 0.70 | 0.46 | 0/1 | 0.94 | 0.24 | 0/1 | Fail |
| Number of years motorcycle license being held by the riders | 10.05 | 13.16 | 0/46 | 16.52 | 15.61 | 0/67 | Fail |
| *Rider-related factors* | | | | | | | |
| Hispanic or Latino[a] | 0.22 | 0.42 | 0/1 | 0.13 | 0.34 | 0/1 | Fail |
| African-American[a] | 0.04 | 0.20 | 0/1 | 0.02 | 0.13 | 0/1 | Fail |
| Height in feet | 5.86 | 0.28 | 4.91/6.75 | 5.83 | 0.26 | 5/6.83 | Pass |
| Age in years | 36.50 | 14.29 | 16/73 | 44.99 | 14.67 | 2/84 | Fail |
| Weight in pounds | 185.95 | 35.39 | 116/362 | 195.26 | 43.02 | 112/480 | Fail |
| Female driver[a] | 0.05 | 0.21 | 0/1 | 0.06 | 0.25 | 0/1 | Pass |
| Driver is college/university graduate[a] | 0.12 | 0.33 | 0/1 | 0.26 | 0.44 | 0/1 | Fail |
| Single driver[a] | 0.31 | 0.46 | 0/1 | 0.44 | 0.50 | 0/1 | Fail |
| Driver is not the owner[a] | 0.11 | 0.31 | 0/1 | 0.05 | 0.21 | 0/1 | Fail |
| Number of children of the driver | 0.91 | 1.30 | 0/5 | 1.26 | 1.38 | 0/6 | Fail |
| Actual speed before event (mph) | 32.84 | 17.24 | 0/90 | 46.35 | 10.90 | 0/85 | Fail |

Notes: N is sample size; $\mu_1$ is mean of crash group; $\mu_2$ is mean of control group; SD is standard deviation; Ho is null hypothesis; (a) indicates indicator variables (1/0); Pass indicates that the null hypothesis cannot be rejected at 95% confidence level; Fail indicates that the null hypothesis can be rejected at 95% confidence level.

The MCCS comprehensive crash causation database also provides detailed information about the rider's apparel and clothing color. Importantly, compared to crash group, control group riders were more frequently wearing motorcycle oriented upper clothing, motorcycle oriented lower clothing, both gloves and shoes that were motorcycle oriented, and retroreflective upper clothing. Other statistics in Table 1 can be interpreted in a similar way[3]. The MCCS data can be used validly for the analysis presented in this study. The results presented in Table 1 help us see if the data (especially the control data) is characterized by erroneous data points. The mean statistics and minimum/maximum values of control data variables in Table 1 are reasonable. For instance, the mean motorcycle driving experience for the crash group riders is 11.52 years, compared to an average experience of 20.48 years for the non-crash group (Table 1). This is intuitive as one would expect that more experienced riders are better skilled and thus less vulnerable to crash involvement. Likewise, compared to the crash group, the proportion of riders with state recognized, entry-level motorcycle course was statistically significantly higher in the non-crash group (see Table 1). Such concordances increase our confidence in the data. Finally, the statistical models based on the data give reasonable results (presented next).

### 3.2. Modeling Results

The detailed empirical analysis presented focuses on analyzing the correlations between motorcycle crash propensity and different explanatory factors shown in Table 1. Different model specifications are presented that either operate at the individual observation level or the matched-triplet level (Figure 1). Also, advanced modeling schemes are explored to fully account for observed and unobserved heterogeneity in motorcycle crash propensity analysis. All the models are derived from a systematic process to include most important variables on basis of statistical significance, specification parsimony, and intuition.

First, fixed parameter logit models (Equation 1) are developed in which the parameter estimates are constrained to be fixed across all observations. As discussed earlier, unobserved heterogeneity and omitted variable bias can be suspected, and in presence of which precise and unbiased correlations cannot be established. Thus, random parameter logit models (Equation 4) are estimated which allow the parameter estimates to vary across the observations. In particular, parameter estimates that either exhibited both statistically significant means and standard deviations, or exhibited only statistically significant standard deviations are treated as random parameters (Fountas and Anastasopoulos, 2017). In the earlier case, both likelihood ratio test and AIC statistics are examined to compare the model treating the specific variable (with only statistically significant standard deviation) as random parameter with the model treating the same variable as fixed parameter (Fountas and Anastasopoulos, 2017). With fixed $\beta$'s and varying $\Upsilon$, the random parameter modeling framework (Equation 4) accounts for the systematic variations (due to unobserved factors) in the associations of explanatory factors (with crash propensity) across the sample population. Next, to account for observed heterogeneity (discussed earlier), random parameter models with heterogeneity-in-means are estimated. The random parameter heterogeneity-in-means approach accounts for both observed and unobserved heterogeneity by allowing the means of random parameters to vary as a function of specific observed factors i.e., explanatory variables shown in Table 1. The goodness-of-fit results of best-fit fixed and random parameter models (Model 1 through Model 4) are presented in Table 2.

As discussed earlier, there exists a matching structure in the data in that the two controls are matched with each focal crash (case) by common matching characteristics. Thus, the empirical framework is extended to also account for both within and between triplet variation and heterogeneity (see methodology section for details). In particular, the goodness-of-fit results of three best-fit models are presented in Table

---

[3] The MCCS data includes a variety of motorcycle types such as, 1) conventional street L1 or L3 vehicles (tank between knees), without modification, 2) conventional street L1 or L3 vehicles (tank between knees), with modifications, 3) Dual purpose, on-road off-road motorcycle, 4) Sport, race replica, 5) Cruiser, 6) Chopper, modified chopper, 7) Touring, 8) Scooter, 9) Tri-cycle, and others. While the statistics for the motorcycle type are not shown in Table 1 for brevity, approximately 41.9%, 23.4%, and 3.1% of the crash-involved motorcycles (N = 351) are sport (race replica motorcycles), cruiser motorcycles, and chopper (or modified chopper) respectively. In addition, 5.4% of the crash-involved motorcycles were scooters and none of the 351 crash-involved motorcycles were tri-cycle.

2, which are: grouped random parameter model (Model 5), grouped random intercept and random parameter model (Model 6), and heterogeneity-in-means grouped random intercept and random parameter model (Model 7).

**TABLE 2: Comparison of Alternative Modeling Frameworks at Individual and Matched-Triplet Levels**

| Goodness of Fit Measures | Models for individual observations (ignoring matched-triplet structure) | | | | Models for matched-triplets (accounting for matched-triplet structure) | | |
|---|---|---|---|---|---|---|---|
| | Model 1 | Model 2 | Model 3 | Model 4 | Model 5 | Model 6 | Model 7 |
| Number of observations | 1053 | 1053 | 1053 | 1053 | 1053 | 1053 | 1053 |
| Number of triplets | --- | --- | --- | --- | 351 | 351 | 351 |
| Degrees of Freedom | 24 | 31 | 32 | 39 | 31 | 32 | 40 |
| Log-likelihood with constant only, Lo | -670.24 | -670.24 | -670.24 | -670.24 | -670.24 | -670.24 | -670.24 |
| Log-likelihood at convergence, Lc | -305.7 | -288.623 | -288.74 | -277.6 | -293.68 | -294.11 | -291.4 |
| Chi-square statistic [2(Lc - Lo)] | 729.08 | 763.234 | 763 | 785.28 | 753.12 | 752.26 | 757.68 |
| AIC | 659.41 | 639.21 | 641.54 | 633.21 | 649.4 | 652.2 | 662.8 |
| Finite Sample AIC | 660.56 | 640.92 | 643.65 | 636.27 | 651.3 | 654.2 | 666.1 |
| McFadden Pseudo $R^2$ | 0.544 | 0.569 | 0.569 | 0.586 | 0.562 | 0.561 | 0.565 |

Notes: AIC Is Akaike Information Criteria.
- Model 1: Fixed parameter logit
- Model 2: Logit model with random parameters across individual observations
- Model 3: Logit model with random intercept and random parameters across individual observations
- Model 4: Logit model with heterogeneity-in-means random parameters across individual observations
- Model 5: Logit model with grouped random parameters across matched-triplets
- Model 6: Logit model with grouped random intercept and random parameters across matched-triplets
- Model 7: Logit model with heterogeneity-in-means grouped random parameters and random intercept across matched-triplets

Apart from Model 1 (fixed parameter logit), the goodness-of-fit results of all competing models suggest that the heterogeneity models which operate at individual observation level clearly outperform their matched-triplet counterparts (Table 2). For instance, logit model with heterogeneity-in-means random parameters across individual observations (Model 1) has the lowest AIC suggesting relatively best fit, followed by logit model with random parameters across individual observations (Model 2), and logit model with random parameters and random intercept across individual observations (Model 3). These results provide compelling evidence that there is no significant within triplet dependence and variation warranting estimation of heterogeneity models operating at matched-triplet level. For brevity, results of models operating at individual observation level are discussed below.

Given the goodness-of-fit statistics (Table 2) discussed above, the results of models operating at individual observation level are discussed in detail from here onwards. A total of 24 explanatory factors are included in the fixed parameter logit model (Model 1), out of which, 17 variables are statistically significant at 95% confidence level (Table 3). Regarding the random parameter logit model (Model 2), a total of seven explanatory factors are found normally distributed random parameters suggesting that their associations with crash propensity vary significantly across crash events. These include: total miles driven prior to crash/interview, one traffic conviction, three traffic convictions, motorcycle oriented lower clothing, female rider, rider is not the motorcycle owner, and speed greater than 50 miles per hour.

The random parameter model (Model 2) resulted in significantly improved fit (lower AIC value compared to Model 1) by accounting for unobserved heterogeneity. However, the goodness-of-fit statistic Table 2 suggest that logit model with heterogeneity-in-means random parameters across individual observations (Model 4) resulted in the best fit with the lowest AIC and Finite Sample AIC of 633.21 and 636.27 respectively, and highest McFadden Pseudo $R^2$ of 0.586. In particular, five of the seven random parameters produced significantly heterogeneity in the means as well (see Table 3). For total miles driven prior to interview/crash, rider with trip originating at work increased the mean making injury crash more likely (Table 3). Contrarily, for total miles driven prior to interview/crash, single rider decreased the mean making injury crash outcome less likely (Table 3). For indicator variable of three traffic convictions, trips originating at home and trips by single rider increased the mean of random parameter, thus making injury crash more likely. Contrarily, the mean of indicator variable for three traffic convictions decreased if the rider had training between 2001-2010 making crash outcomes less likely. Regarding the heterogeneous effects of indicator variable for one traffic conviction due to unobserved heterogeneity, the random parameter also exhibited observed heterogeneity where the mean increased as function of observed explanatory variable (i.e., single rider). Interestingly, evaluated at the mean proportion of Hispanic or Latino riders in the data, events involving female riders exhibited smaller means suggesting lower probability of crash outcomes. Finally, if the rider speed is greater than 50 miles per hour, events in which rider consumed alcohol or multiple drugs significantly increase the mean of the random parameter making crash a probable outcome.

In addition to the goodness-of-fit statistics, the above conceptual findings reveal the significant potential of heterogeneity based models (both observed and unobserved) in providing richer insights into the phenomenon under investigation. To better interpret the estimation results of the best-fit heterogeneity-in-means random parameter logit model, the percent relative risk estimates are provided in Table 4. The relative risks in Table 4 are estimated in terms of odds ratio which is the exponent of the parameter estimates, **β**, and are calculated as 100 x (exp(**β**)-1). For each one unit increase (and switch from 0 to 1 for a dummy variable) in the value of explanatory factor, the relative risks show the percent increase (decrease) in the odds of riders getting involved in a crash. To demonstrate the contrast in relative risks generated by different methodological alternatives, the relative risks for fixed parameter and random parameter logit models are also provided in Table 4. Finally, to help conceptualize the distribution effects of random-held parameters, key distributional statistics are provided in Table 5.

From a conceptual perspective, the rigorous statistical analysis was conducted to achieve four objectives. The first objective was to investigate how motorcyclist apparel and visual conspicuity influences their likelihood of being involved in a crash. Second, how motorcyclist's training (especially with relevance to the recentness of the training), education programs, and type of helmet coverage are related to crash outcomes. Third, how previous traffic convictions of the rider are related to his/her crash propensity. This is a complicated issue as previous traffic convictions can produce deterrence and moral hazard effects as well as can indicate consistently risky riders. Fourth, how behavioral factors (fatigue, speeding, alcohol, multiple drugs intake), trip-related factors, and rider-related factors (age, experience, gender, etc.) relate to the chances of getting involved into a crash. As the observed associations of aforementioned factors can be complex and heterogeneous, the possibility of systematic variations in the associations due to observed as well as unobserved factors is explicitly considered in the estimation.

Interesting findings regarding the correlations between the motorcycle crash propensity and key explanatory variables pertaining to aforementioned objectives are discussed next.

**TABLE 3: Estimation Results for Fixed Parameter Logit, Random Parameter Logit, and Heterogeneity-in-Means Random Parameter Logit**

| Variables | Fixed Parameter Logit (Model 1) | | Random Parameter Logit (Model 2) | | Random Parameter Logit - Heterogeneity in Means (Model 4) | |
|---|---|---|---|---|---|---|
| | β | t-stat | β | t-stat | β | t-stat |
| ***Random Parameters*** | | | | | | |
| Total miles driven prior to crash/interview | -0.003 | -0.63 | -0.026 | -3.31 | -0.008 | -1.02 |
| *scale parameter* | --- | --- | *0.064* | *6.5* | *0.051* | *5.49* |
| One traffic conviction | 0.445 | 1.78 | 0.202 | 0.86 | -0.309 | -0.92 |
| *scale parameter* | --- | --- | *1.596* | *5.31* | *1.65* | *5.24* |
| Three traffic convictions | 0.484 | 1.01 | -4.541 | -2.29 | -16.99 | -2.54 |
| *scale parameter* | --- | --- | *19.125* | *3.5* | *33.8* | *2.98* |
| Lower clothing motorcycle oriented | -1.497 | -4.35 | -4.519 | -4.95 | -6.5 | -4.76 |
| *scale parameter* | --- | --- | *4.988* | *5.45* | *7.24* | *5.41* |
| Female rider | 0.41 | 0.93 | -0.066 | -0.15 | 0.39 | 0.73 |
| *scale parameter* | --- | --- | *2.286* | *3.51* | *1.71* | *2.74* |
| Rider is not the owner | -0.741 | -1.57 | -0.872 | -1.68 | -1.16 | -2.15 |
| *scale parameter* | --- | --- | *2.439* | *3.39* | *2.89* | *3.75* |
| Speed greater than 50 mph | -1.415 | -3.96 | -2.687 | -4.55 | -2.98 | -4.6 |
| *scale parameter* | --- | --- | *3.417* | *4.93* | *3.49* | *4.77* |
| ***Heterogeneity in the Means of Random Parameter (Total miles driven prior to crash/interview)*** | | | | | | |
| *Origin: Work* | --- | --- | --- | --- | *0.037* | *1.87* |
| *Single rider* | --- | --- | --- | --- | *-0.044* | *-3.38* |
| ***Heterogeneity in the Means of Random Parameter (Three traffic convictions)*** | | | | | | |
| *Origin: Work* | --- | --- | --- | --- | *8.249* | *1.99* |
| *Single rider* | --- | --- | --- | --- | *13.08* | *2.74* |
| *Training between 2001-2010* | --- | --- | --- | --- | *-9.037* | *-2.18* |
| ***Heterogeneity in the Means of Random Parameter (One traffic conviction)*** | | | | | | |
| *Single rider* | --- | --- | --- | --- | *0.852* | *1.99* |
| ***Heterogeneity in the Means of Random Parameter (Female rider)*** | | | | | | |
| *Hispanic or Latino rider* | --- | --- | --- | --- | *-1.105* | *-1.56* |
| ***Heterogeneity in the Means of Random Parameter (Speed greater than 50 mph)*** | | | | | | |
| *Alcohol and multiple drugs* | --- | --- | --- | --- | *2.628* | *2.57* |

Notes: **β** is parameter estimate; (---) indicates not applicable;

**TABLE 3: Estimation Results for Fixed Parameter Logit, Random Parameter Logit, and Heterogeneity-in-Means Random Parameter Logit**
*(Continued)*

| Variables | Fixed Parameter Logit (Model 1) | | Random Parameter Logit (Model 2) | | Random Parameter Logit - Heterogeneity in Means (Model 4) | |
|---|---|---|---|---|---|---|
| | β | t-stat | β | t-stat | β | t-stat |
| *Fixed Parameters* | | | | | | |
| Constant | 3.32 | 4.8 | 4.41 | 6.21 | 4.62 | 6.1 |
| Origin: Home | -2.46 | -7.97 | -3.00 | -8.04 | -3.08 | -7.86 |
| Origin: Work | -1.24 | -3.12 | -1.68 | -4.05 | -2.09 | -4.25 |
| Destination: Friend/relative place | 1.36 | 3.66 | 1.49 | 4.06 | 1.55 | 4.32 |
| 5 hours or less sleep | 0.92 | 2.43 | 1.07 | 3.31 | 1.09 | 3.36 |
| Road used daily | 0.46 | 2.02 | 0.46 | 2.09 | 0.5 | 2.26 |
| Road used once per month | -0.85 | -2.27 | -1.08 | -2.8 | -1.06 | -2.6 |
| Helmet coverage type 1 (Partial coverage) | -0.76 | -2.71 | -0.73 | -2.63 | -0.68 | -2.45 |
| Training between 2001-2010 | -1.05 | -3.78 | -1.21 | -4.44 | -1.15 | -4.13 |
| Training between 2011- 2015 | -1.33 | -4.14 | -1.48 | -4.88 | -1.43 | -4.68 |
| Two traffic convictions | 0.828 | 2.77 | 0.82 | 2.76 | 0.85 | 2.87 |
| Upper body clothing color: Red | 1.131 | 2.47 | 1.27 | 2.8 | 1.38 | 3.04 |
| Motorcycle license being held by the rider for 30 or more years | -0.481 | -1.38 | -0.44 | -1.27 | -0.36 | -1.01 |
| Hispanic or Latino driver | 0.544 | 2.06 | 0.70 | 2.67 | 0.77 | 2.71 |
| Rider age in years | -0.029 | -3.09 | -0.03 | -3.44 | -0.04 | -3.79 |
| Rider weight in pounds | -0.004 | -1.8 | -0.01 | -2.66 | -0.007 | -2.62 |
| Rider is college/university graduate | -0.295 | -1.15 | -0.29 | -1.14 | -0.28 | -1.06 |

Notes: β is parameter estimate.

## 4. DISCUSSION

The results and findings discussed here refer to the random parameter logit model with heterogeneity-in-means (Model 4) given its best fit among all competing models (Table 3). Overall, the crash propensity models quantify the associations between a wide variety of factors and the likelihood of an injury crash. In the discussion of results, the word "crash" is used to refer to an injury crash.

### 4.1. Rider's Apparel and Visual Conspicuity

Several factors related to rider's apparel, conspicuity, and history of traffic convictions in last 5 years are statistically significantly correlated with the likelihood of a crash. While the association between motorcycle rider conspicuity and injury risk have been studied (Wells et al., 2004), little to no information exists about how rider's apparel may be correlated with his/her risk of crash involvement. Our analysis shows that riders whose lower clothing is motorcycle oriented have a lesser crash involvement propensity (risk estimates of -99.8 percentage points). Despite heterogeneity in the effects, for a significant portion of the sample, i.e., 81.5%, the association of lower clothing being motorcycle oriented with crash propensity is negative. This suggests that appropriate apparel can reduce a motorcyclist's crash likelihood. In addition, motorcycle rider conspicuity, i.e., detectability and visibility on road, is regarded as a "high-priority" key risk factor in the recent US DOT's National Agenda for Motorcycle Safety (NHTSA, 2013). One of the key finding of the famous Hurt Report was that "motorcycle riders with high conspicuity were less likely to have their right-of-way violated by other vehicles." (Hurt et al., 1981). After adjusting for observed and unobserved heterogeneity, our analysis suggests that riders with "red" upper body clothing are associated with a 297% increase in the odds of crash involvement. This is intuitive as dark color upper clothing may reduce the motorcycle rider's conspicuity, i.e., rider's visibility or detectability on road. Also, past research by Wells et al. (2004) (Wells et al., 2004) found that dark color helmets and dark waist up clothing (lower conspicuity) are typically associated with higher likelihood of injury involved crashes (Wells et al., 2004).

### 4.2. Type of Helmet Coverage and Training Programs

The statistical models also quantify the associations between type of helmet coverage, year of motorcycle training, and crash propensity. It is found that helmets with partial coverage (US DOT compliant helmets with partial coverage, least intrusive covering only the top half of the cranium) are associated with lower likelihood of crash involvement. The is somewhat intuitive; as partial coverage helmets may not interfere with the rider's hearing and vision capabilities, and thus may decrease the probability of a crash. Likewise, if a rider received training in recent years (training between 2001-2010, and training between 2011-2015), their likelihood of crash involvement is lower by 69% and 77% respectively (Table 5). This finding is also intuitive as motorcycle rider training programs in recent years have significantly improved owing to the national and statewide efforts for improving motorcycle safety (NHTSA, 2013).

All of above findings relate to "policy-sensitive" and "preventable" key risk factors. For instance, by encouraging riders to increase their conspicuity and/or by using motorcycle oriented rider clothing, a reduction in motorcycle crashes can be achieved. Likewise, awareness programs aimed at encouraging motorcycle riders to participate in formal training programs can also help alleviate the crash risks.

### 4.3. Previous Traffic Convictions

The variables related to the number of traffic convictions in past five years are statistically significantly related to the likelihood of crash. Broadly, one would expect that riders with more traffic convictions in past (likely to be unsafe riders) would be more likely to get involved in a crash. In the fixed parameter logit model (Model 2 in Table 3), this is reflected in the positive and statistically significant parameter estimate for two traffic convictions, and positive but statistically not significant estimates for one and three traffic convictions (Table 3). However, both indicator variables for one and three traffic convictions became statistically significant in the random parameter heterogeneity-in-means counterpart. Importantly, the random parameter model suggests that riders with three traffic conviction in past are on-average less likely to get involved in a crash (negative and positive association for 69.24% and 30.76% of observations respectively) (Table 5). This finding related to heterogeneity is important and has at least two dimensions, 1) riders who have more traffic convictions in past may react to this penalty by driving (or riding) more cautiously in the future, and 2) some riders who have more traffic convictions in past may be consistent risky riders and may exhibit higher likelihood of crash involvement given their risky riding behaviors.

Owing to the need for confirming or discarding the directional consistency in the associations between traffic convictions and crash propensity, we can discuss and compare two important dimensions owing to heterogeneity in light of other empirical studies. The principles of deterrence theory and moral hazard is relevant to the first dimension, and posits that a strong penalty (in terms of demerit points, fines, etc.) can be more effective in reducing or preventing recidivism. If a driver (or rider) has more demerit points on their license, they may react to this penalty by driving (or riding) more cautiously in the future. For instance, Cooper (1990) found support for the hypothesis that drivers who have fewer accidents may exhibit a worse safety record (based on driving exposure with non-crash convictions as a surrogate measure of their safety record) (Cooper, 1990). In particular, the study observed that while elderly drivers have fewer accidents than younger drivers, their safety record is worse and in fact are more often found to be at fault in accidents. Along these lines, a recent study identified the "deterrent effects" of Demerit Points System (DPS) by using variables related to accumulated demerit points (Lee et al., 2018). The study by Lee et al. (2018) quantified the connection between relative violation hazards and accumulated demerit points while controlling for drivers who had no license sanctions, who had license suspension experience, and those who had license revocation experience (Lee et al., 2018). The study concluded, both for suspension and revocation limits, that demerit point accumulation increased the traffic law compliance duration of previous traffic infringers (Lee et al., 2018). Several other studies also conclude evidence in support of moral hazard, i.e., drivers who accumulate demerit points (more traffic convictions) tend to be more careful owing to the risk of losing their driving licenses. For details, see (Dionne et al., 2011, Lee et al., 2018, Basili et al., 2015).

On the other hand, previous studies have also found a positive association between crash risk and past traffic convictions. In the context of motor vehicle driver injury, Blows et al. (2005) found that drivers with higher numbers of convictions in the past 12 months are more likely to exhibit a higher risk of injury (Blows et al., 2005). Using random selection of license holders as controls, Rajalin (1994) conducted a population based case-control study in Finland that examined the connection between risky driving and involvement in fatal accidents (Rajalin, 1994). After controlling for age and an estimate of miles driven by the driver annually, the study found that fatally crashed male drivers were statistically significantly more likely to have traffic convictions (offences) on their driving records (Rajalin, 1994). Likewise, a literature review of explanatory factors correlated with increased crash risk found that prior traffic convictions (offences) are good predictors of crash risk[4] (Peck, 1993).

## 4.4. Trip and Exposure-Related Factors

Regarding trip-related factors, trips originating either at home (indicator variable for Origin: Home) or at work (indicator variable for Origin: Work) are statistically significantly negatively associated with likelihood of crash. For instance, a trip originated at home and work place decreases the odds of crash by 95.4 and 97.6 percentage points (Table 4). Contrarily, trips where the destination is friend/relative place are positively associated with the likelihood of a crash. Regarding exposure-related factors, the variable related to rider holding a motorcycle license for more than 30 years, and the variable related to miles ridden before the crash/interview are on-average negatively associated with likelihood of a crash. However, the parameter estimates for variable related to total miles driven before crash/interview exhibit significant heterogeneity. For instance, the association is negative for 56.23% of observations and positive for 43.77% of the observations. In addition, the associations also exhibited significant heterogeneity in the means, with mean effects varying as a function of trip origin being work place and single rider, as discussed earlier (see Table 3). Note that this variable was statistically not significant in the fixed-parameter counterpart (Model 2 in Table 3).

---

[4] Keeping in view the results of previous studies and theoretical basis, the directional inconsistency in the associations of traffic convictions with crash propensity cannot be discarded. It is highly likely that the heterogeneity in magnitude and direction of the correlations between traffic conviction related variables and crash propensity is tracking the two dimensions (discussed above). As can be seen, the rigorous methodological framework adopted in this study helps address this issue by accounting for the heterogeneous correlations between key risk factors and motorcyclist crash propensity (Table 5).

**TABLE 4: Relative Risk Estimates for Motorcycle Crash Propensity***

| Variables | Fixed Parameter Logit (Model 1) | | Random Parameter Logit (Model 2) | | Random Parameter Logit - Heterogeneity in Means (Model 4) | |
|---|---|---|---|---|---|---|
| | Direction of association | % change in crash risk | Direction of association | % change in crash risk | Direction of association | % change in crash risk |
| **Exposure-related factors** | | | | | | |
| Total miles driven prior to event | ↓ | -0.300 | [↓] | -2.57 | [↓]a | -0.80 |
| **Number of traffic convictions in last 5 years** | | | | | | |
| One traffic conviction | ↑ | 56.05 | [↑] | 22.38 | [↓]a | -26.58 |
| Two traffic convictions | ↑ | 128.87 | ↑ | 127.28 | ↑ | 133.96 |
| Three traffic convictions | ↑ | 62.26 | [↓] | -98.93 | [↓]a | -101.00 |
| **Clothing color** | | | | | | |
| Lower clothing motorcycle oriented | ↓ | -77.62 | [↓] | -98.91 | [↓] | -99.85 |
| Upper body clothing color: Red | ↑ | 209.88 | ↑ | 254.31 | ↑ | 297.49 |
| **Rider-related factors** | | | | | | |
| Motorcycle license being held by the rider for 30 or more years | ↓ | -38.18 | ↓ | -35.85 | ↓ | -30.23 |
| 5 hours or less sleep | ↑ | 150.93 | ↑ | 191.54 | ↑ | 197.43 |
| Female rider | ↑ | 50.68 | [↓] | -6.39 | [↑]a | 47.70 |
| Rider is not the owner | ↓ | -52.34 | [↓] | -58.19 | [↓] | -68.65 |
| Hispanic or Latino rider | ↑ | 72.29 | ↑ | 101.78 | ↑ | 115.98 |
| Rider age in years | ↓ | -2.86 | ↓ | -2.96 | ↓ | -3.92 |
| Rider weight in pounds | ↓ | -0.399 | ↓ | -0.60 | ↓ | -0.70 |
| Rider is college/university graduate | ↓ | -25.55 | ↓ | -25.32 | ↓ | -24.42 |
| **Trip-related factors** | | | | | | |
| Origin: Home | ↓ | -91.46 | ↓ | -95.04 | ↓ | -95.40 |
| Origin: Work | ↓ | -71.06 | ↓ | -81.40 | ↓ | -87.63 |
| Destination: Friend/relative place | ↑ | 289.62 | ↑ | 341.94 | ↑ | 371.15 |
| **Frequency of road use** | | | | | | |
| Road used daily | ↑ | 58.41 | ↑ | 57.93 | ↑ | 64.87 |
| Road used once per month | ↓ | -57.26 | ↓ | -66.14 | ↓ | -65.35 |
| **Type of helmet coverage** | | | | | | |
| Helmet coverage type 1 (Partial coverage) | ↓ | -53.23 | ↓ | -51.81 | ↓ | -49.34 |
| **Year of training** | | | | | | |
| Training between 2001-2010 | ↓ | -65.01 | ↓ | -70.09 | ↓ | -68.34 |
| Training between 2011- 2015 | ↓ | -73.55 | ↓ | -77.26 | ↓ | -76.07 |
| **Speed before crash/interview** | | | | | | |
| Speed greater than 50 mph | ↓ | -75.71 | [↓] | -93.19 | [↓]a | -94.92 |

Notes: (*) Parentheses indicate mixed effects for the random-held parameters; (a) indicates random parameters with heterogeneity-in-means.

**TABLE 5: Distribution Effects of the Random Parameters in Random Parameter Logit and Random Parameter Logit with Heterogeneity-in-the-Means (Models 2 and 4).**

| Variables | Random Parameter Logit (Model 2) | | | | Random Parameter Logit-Heterogeneity in Means (Model 4) | | | |
|---|---|---|---|---|---|---|---|---|
| | μ | SD | Above zero | Below zero | μ | SD | Above zero | Below zero |
| **Exposure-related factors** | | | | | | | | |
| Total miles driven prior to event | -0.026 | 0.064 | 34.23% | 65.77% | -0.008 | 0.051 | 43.77% | 56.23% |
| **Number of traffic convictions in last 5 years** | | | | | | | | |
| One traffic conviction | 0.202 | 1.596 | 55.04% | 44.96% | -0.309 | 1.65 | 42.57% | 57.43% |
| Three traffic convictions | -4.541 | 19.125 | 40.62% | 59.38% | -16.99 | 33.8 | 30.76% | 69.24% |
| **Clothing color** | | | | | | | | |
| Lower clothing motorcycle oriented | -4.519 | 4.988 | 18.25% | 81.75% | -6.5 | 7.24 | 18.46% | 81.54% |
| **Rider-related factors** | | | | | | | | |
| Female rider | -0.066 | 2.286 | 48.85% | 51.15% | 0.39 | 1.71 | 59.07% | 40.93% |
| Rider is not the owner | -0.872 | 2.439 | 36.035 | 63.67% | -1.16 | 2.89 | 34.41% | 65.59% |
| Speed greater than 50 mph | -2.687 | 3.417 | 21.58% | 78.42% | -2.98 | 3.49 | 19.66% | 80.34% |

Notes: μ is the mean of random parameter; SD is the standard deviation of random parameter.

## 4.5. Behavioral and Rider-Related Factors

The estimation results shed light on the associations between several important rider behavioral factors such as fatigue, speeding, alcohol, and multiple drugs intake with crash propensity. The results show that if the rider's speed is greater than 50 mph, the likelihood of crash involvement is low, albeit with significant heterogeneity in magnitude as well as direction of associations (see Table 3 and 5). This may be so as high speeds often imply trips on higher classification roadways. Note that trips on roadways with higher classification (such as arterials and freeways) may have small crash frequency, but significantly higher injury outcomes, given a crash (Quddus et al., 2002). Also, the associations exhibit significant heterogeneity in the means, with mean correlation of speed varying as a function of alcohol and multiple drugs involvement. That is, if a rider's speed is greater than 50 miles per hour, events in which rider also consumed alcohol or multiple drugs significantly increase the mean of the random parameter making crash a more probable outcome. This finding is important and suggests development of countermeasures aimed at reducing riding under the influence, as the consequences of alcohol or multiple drugs use can be severe especially at high speeds. Finally, if the rider had 5 hours or less sleep (surrogate of drowsy-riding and/or fatigue) before the crash/interview, his/her odds of getting involved into a crash increased by approximately 200 percentage points (Table 5). Note the significant contrast in the relative risk estimates of fixed and random parameter counterparts (Table 5). This points to the development of programs targeted at reducing motorcycle riding while being sleepy (or fatigued). The models also quantify the heterogeneous correlations between driver-related factors, frequency of road use, education, gender, and crash propensity (see Table 3 for details).

Related to rider-related factors, the estimation results quantify the correlations between age, gender, rider race, vehicle ownership, and crash propensity. The results suggest that an increase in rider age is positively correlated with crash propensity (Table 3). In particular, a one-unit increase in rider age decreases the risk of crash by 3.92% in random parameter heterogeneity in means logit, compared to a 2.86% decrease in risk shown in fixed parameter logit model (Table 4). This finding is intuitive as riders with greater age may have more riding experience, and is in agreement with previous literature (Lin et al., 2003, Cooper, 1990). Compared to male riders, female riders have a higher chance of getting involved in a crash albeit significant heterogeneity in magnitude as well as direction of associations. For instance, the correlation between female riders and crash propensity is positive for 59.07% of the cases, whereas negative for 40.93% of the cases (see the distribution effects in Table 5). Compared to riders with other origins/race, the results show that Hispanic or Latino riders are more likely to get involved in a crash. That is, if the rider is Hispanic or Latino, his/her odds of getting involved into a crash increases by 115.98%, compared to only 72.29% increase indicated by fixed parameter logit model (Table 4). Finally, compared to riders riding their own motorcycles, the chance of a crash reduce by 68.65% for riders who do not own the motorcycles being ridden (Table 4). However, heterogeneity in the direction of association exists as the relationship is negative for 65.59% of the riders and positive for the rest (see distributional effects in Table 5).

## 5. LIMITATIONS

The present study is based on a sample of 1053 events in Orange County, California, out of which 351 were identified as injury crash events. This study uses MCCS data, which is the most comprehensive national effort to-date. However, the results should be interpreted with caution due to the limited sample size. Also, the data are collected in Orange County, California and may not be representative of motorcyclists in other areas of California and the U.S. Related to general factors characterizing the reference area, the population of Orange County, California is 3,172,532, with 14% of the residents being 65 years and over (15.2% in the entire U.S. and 13.6% in California), and with 5.1% of the residents having a disability under the age of 65 years (6.8% in California) *(https://www.census.gov/quickfacts/fact/table/orangecountycalifornia/PST045216)*. Moreover, 84.5 percent of the total county population are high school graduates or possess higher qualifications, compared to 87% in the entire U.S. and 82.1% in California. Regarding living arrangements, Orange County has a total of 1,017,012 households are located in Orange County, with an average of 3.04 persons per household (2012-2016) (2.95 in California), and with a median household income of $78,145 (in 2016 dollars) from 2012-2016. Finally, in terms of geography, the total land area in square miles of Orange County is 790.57 miles (2010) with approximately 3807 residents per square mile (2010). These statistics may allow readers to establish similarity patterns with other locations potentially presenting similar features in these regards. While Orange County may be unique in certain respects, the statistics presented above indicate that it is fairly representative of California and the U.S. In addition, the behaviors leading to crashes observed in the MCCS are not uncommon and as such can be regarded as reasonably representative of the general behaviors of motorcyclists.

The matched case-control design allows efficient analysis of rare diseases (injury crashes) while controlling for the exposure of the population under study (i.e., motorcyclists) to the outcome of interest (i.e., motorcyclist crash). However, a frequently reported disadvantage of matched case-control studies is the retrospective nature. That is, the study framework looks backwards and investigates exposure to crash risk. As such, retrospective studies may be exposed to errors related to confounding and bias. However, as the investigations in this study are performed in field by trained experts, and do not build on investigator's memory per se, the extent of confounding and recall bias is likely small.

## 6. CONCLUSIONS

The study has focused on how different "policy-sensitive" factors are associated with a motorcyclist's risk of involvement in an injury crash while controlling for rider-specific, psycho-physiological, and other observed/unobserved factors. The analysis used data from a comprehensive matched case-control design collected through the US DOT Federal Highway Administration's Motorcycle Crash Causation Study (MCCS). To help account for the exposure of the population under study, we analyzed the reported 351 motorcycle-involved injury crashes vis-à-vis 702 similarly-at-risk matched controls (motorcyclists not involved in crashes).

Unlike traditional conditional estimation of relative risks, heterogeneity based case-control analysis is presented to quantify the relative risks of different "policy-sensitive" variables. By doing so, the possibilities of both within and between matched case-control variations due to observed and unobserved factors are considered. We estimated fixed parameter logit, random parameter logit, and heterogeneity-in-means random parameter logit models at the individual observation level as well as at the matched-triplet level. From an empirical standpoint, the findings suggest that heterogeneity based models operating at the individual observation level clearly outperform their matched triplet counterparts. Random parameter logit model with heterogeneity-in-means provided the best-fit, and shows that the correlations between key risk factors and crash propensity exhibit significant observed and unobserved heterogeneity.

Several important findings surfaced from the empirical analysis. Regarding motorcycle helmet coverage, analysis shows that riders with partial helmet coverage (US DOT compliant helmets with partial coverage, least intrusive covering only the top half of the cranium) have a lower risk of injury crash involvement. Our analysis suggests that lack of motorcycle rider conspicuity captured by dark (red) upper body clothing is associated with a significantly higher injury crash risk. Related to the appropriateness of motorcycle rider's clothing, if the rider's lower clothing is motorcycle-oriented (e.g., cannot easily get stuck in the machinery or wheels), then the odds of injury crash involvement are lower. All of these findings relate to "policy-sensitive" and "preventable" key risk factors. The conclusions are that reductions in motorcycle injury crashes are possible by encouraging helmet usage, increasing rider conspicuity and/or by using motorcycle oriented lower clothing.

Related to the effectiveness of formal motorcycle training programs, behavioral factors such as speeding, fatigue or likely drowsy riding, alcohol/drugs intake, the conclusions are that ongoing participation in motorcycle training programs (perhaps through refresher courses for experienced motorcyclists) and prevention of sleep-deprived/fatigued riding and riding under the influence of alcohol and drugs (especially at high speeds) can potentially improve motorcyclist safety. A future study could quantify the number of lives that would be saved or injuries prevented annually if helmet use was to become ubiquitous or if certain types of helmets were worn properly (a more nuanced finding from this paper). This study quantifies the effectiveness of a range of motorcycle-specific interventions. There is a need to estimate the population that will benefit from these interventions and how these interventions can be implemented at a broader scale. The empirical work presented in this paper lays the foundation for simulation modeling to provide guidance and recommendations to state and local agencies on how to prioritize their road safety policies and programs to reduce motorcycle crash risks.

## 7. ACKNOWLEDGEMENT

This paper is based on work supported by the US Department of Transportation through the Collaborative Sciences Center for Road Safety (CSCRS), a consortium led by The University of North Carolina at Chapel Hill (UNC) in partnership with The University of Tennessee. The title of the on-going research project (R-20) is "Investigating the Vulnerability of Motorcyclists to Crashes and Injury." The data for this study were provided by the Federal Highway Administration (FHWA). The assistance of Ms. Carol Tan and Mr. Yusuf Mohamedshah is highly appreciated. The paper greatly benefited from the detailed discussions with Dr. Clyde Schechter at Albert Einstein College of Medicine and his valuable advice on the study design and the empirical methods. Finally, the study has significantly benefited from the valuable input of Dr. Arthur Goodwin at the UNC Highway Safety Research Center. The opinions, findings, and conclusions or recommendations expressed in this paper are those of the authors and do not necessarily reflect the views of the FHWA/US DOT. Any remaining errors are ours.

# 8. REFERENCES

Atzmon, G., Schechter, C., Greiner, W., Davidson, D., Rennert, G. & Barzilai, N. 2004. Clinical phenotype of families with longevity. *Journal of the American Geriatrics Society,* 52**,** 274-277.

Barzilai, N., Atzmon, G., Schechter, C., Schaefer, E. J., Cupples, A. L., Lipton, R., Cheng, S. & Shuldiner, A. R. 2003. Unique lipoprotein phenotype and genotype associated with exceptional longevity. *Jama,* 290**,** 2030-2040.

Basili, M., Belloc, F. & Nicita, A. 2015. Group attitude and hybrid sanctions: Micro-econometric evidence from traffic law. *Transportation Research Part A: Policy and Practice,* 78**,** 325-336.

Behnood, A. & Mannering, F. 2017. The effect of passengers on driver-injury severities in single-vehicle crashes: A random parameters heterogeneity-in-means approach. *Analytic Methods in Accident Research,* 14**,** 41-53.

Bhat, C. R. 2003. Simulation estimation of mixed discrete choice models using randomized and scrambled Halton sequences. *Transportation Research Part B: Methodological,* 37**,** 837-855.

Blackman, R. A. & Haworth, N. L. 2013. Comparison of moped, scooter and motorcycle crash risk and crash severity. *Accident Analysis & Prevention,* 57**,** 1-9.

Blows, S., Ameratunga, S., Ivers, R. Q., Lo, S. K. & Norton, R. 2005. Risky driving habits and motor vehicle driver injury. *Accident Analysis & Prevention,* 37**,** 619-624.

Boakye, K., Wali, B., Khattak, A. & Nambisan, S. 2018. Are Enforcement Strategies Effective in Increasing Nighttime Seat Belt Use? Evidence from a Large-Scale Before–After Observational Study. *Presented at Transportation Research Board 97th Annual Meeting.* Washington DC, 2018.

Brandt, M.-M., Ahrns, K. S., Corpron, C., Franklin, G. A. & Wahl, W. L. 2002. Hospital cost is reduced by motorcycle helmet use. *Journal of Trauma and Acute Care Surgery,* 53**,** 469-471.

Broughton, P., Fuller, R., Stradling, S., Gormley, M., Kinnear, N., O'dolan, C. & Hannigan, B. 2009. Conditions for speeding behaviour: a comparison of car drivers and powered two wheeled riders. *Transportation research part F: traffic psychology and behaviour,* 12**,** 417-427.

Cheng, A. S., Liu, K. P. & Tulliani, N. 2015. Relationship between driving-violation behaviours and risk perception in motorcycle accidents. *Hong Kong Journal of Occupational Therapy,* 25**,** 32-38.

Chin, H. C. & Quddus, M. A. 2003. Modeling count data with excess zeroes: an empirical application to traffic accidents. *Sociological methods & research,* 32**,** 90-116.

Cooper, P. J. 1990. Differences in accident characteristics among elderly drivers and between elderly and middle-aged drivers. *Accident Analysis & Prevention,* 22**,** 499-508.

Dionne, G., Pinquet, J., Maurice, M. & Vanasse, C. 2011. Incentive mechanisms for safe driving: a comparative analysis with dynamic data. *The review of Economics and Statistics,* 93**,** 218-227.

Elliott, M. A., Baughan, C. J. & Sexton, B. F. 2007. Errors and violations in relation to motorcyclists' crash risk. *Accident Analysis & Prevention,* 39**,** 491-499.

FHWA. 2017a. *Motorcycle Crash Causation Study (MCCS). US Department of Transportation.* <https://www.fhwa.dot.gov/research/tfhrc/projects/safety/motorcycles/mccs/> [Online]. US Department of Transportation.

FHWA 2017b. Motorcycle Crash Causation Study: Final Report. Cooperative Agreement No. DTFH61-06-H-00034. Federal Highway Administration, US Department of Transportation.

Fountas, G. & Anastasopoulos, P. C. 2017. A random thresholds random parameters hierarchical ordered probit analysis of highway accident injury-severities. *Analytic methods in accident research,* 15**,** 1-16.

Fuller, R., Hannigan, B., Bates, H., Gormley, M., Stradling, S., Broughton, P., Kinnear, N. & O'Dolan, C. 2008. Understanding inappropriate high speed: a qualitative analysis. *Department for Transport*.

Haque, M. M., Chin, H. C. & Huang, H. 2010. Applying Bayesian hierarchical models to examine motorcycle crashes at signalized intersections. *Accident Analysis & Prevention,* 42**,** 203-212.

Hartman, E., Frankena, K., Vrielink, H. H. O., Nielen, M., Metz, J. H. & Huirne, R. B. 2004. Risk factors associated with sick leave due to work-related injuries in Dutch farmers: an exploratory case-control study. *Safety science,* 42**,** 807-823.
Hurt, H. H., Ouellet, J. V. & Thom, D. R. 1981. *Motorcycle accident cause factors and identification of countermeasures*, The Administration.
Keall, M. D. & Newstead, S. 2012. Analysis of factors that increase motorcycle rider risk compared to car driver risk. *Accident Analysis & Prevention,* 49**,** 23-29.
Khan, W. A. & Khattak, A. J. 2018. Injury Severity of Truck Drivers in Crashes at Highway–Rail Grade Crossings in the United States. *Presented at the Transportation Research Board 97th Annual Meeting, Washington DC, 2018.*
Khattak, A. J. & Wali, B. 2017. Analysis of volatility in driving regimes extracted from basic safety messages transmitted between connected vehicles. *Transportation research part C: emerging technologies,* 84**,** 48-73.
Lee, J., Park, B.-J. & Lee, C. 2018. Deterrent effects of demerit points and license sanctions on drivers' traffic law violations using a proportional hazard model. *Accident Analysis & Prevention,* 113**,** 279-286.
Li, X., Khattak, A. J. & Wali, B. 2017. Role of Multiagency Response and On-Scene Times in Large-Scale Traffic Incidents. *Transportation Research Record: Journal of the Transportation Research Board***,** 39-48.
Lin, M.-R., Chang, S.-H., Pai, L. & Keyl, P. M. 2003. A longitudinal study of risk factors for motorcycle crashes among junior college students in Taiwan. *Accident Analysis & Prevention,* 35**,** 243-252.
Liu, J., Khattak, A. J. & Wali, B. 2017. Do safety performance functions used for predicting crash frequency vary across space? Applying geographically weighted regressions to account for spatial heterogeneity. *Accident Analysis & Prevention,* 109**,** 132-142.
Mannering, F. L. & Bhat, C. R. 2014. Analytic methods in accident research: Methodological frontier and future directions. *Analytic methods in accident research,* 1**,** 1-22.
Mannering, F. L., Shankar, V. & Bhat, C. R. 2016. Unobserved heterogeneity and the statistical analysis of highway accident data. *Analytic methods in accident research,* 11**,** 1-16.
Meuleners, L., Fraser, M. L., Govorko, M. H. & Stevenson, M. R. 2015. Determinants of the occupational environment and heavy vehicle crashes in Western Australia: A case–control study. *Accident Analysis & Prevention*.
Moskal, A., Martin, J.-L. & Laumon, B. 2012. Risk factors for injury accidents among moped and motorcycle riders. *Accident Analysis & Prevention,* 49**,** 5-11.
NHTSA 2010. Motorcycle Crash Causes And Outcomes: Pilot Study. US Department of Transportation: National Highway Traffic Safety Administration. URL: http://www.nhtsa.gov/staticfiles/nti/pdf/811280.pdf.
NHTSA. 2013. *National Agenda for Motorcycle Safety* [Online]. Washington DC. Available: https://one.nhtsa.gov/people/injury/pedbimot/motorcycle/00-NHT-212-motorcycle/toc.html.
NHTSA 2016. 2015 motor vehicle crashes: Overview. Fatality Analysis Reporting System Encyclopedia, National Highway Traffic Safety Administration. URL: https://crashstats.nhtsa.dot.gov/Api/Public/Publication/812384. *Traffic safety facts research note.*
NHTSA 2017. Motorcycle Crash Causation Study. *Federal Highway Administration, U.S. Department of Transportation.*
Peck, R. C. 1993. The identification of multiple accident correlates in high risk drivers with specific emphasis on the role of age, experience and prior traffic violation frequency. *Alcohol, Drugs & Driving*.
Preusser, D. F., Williams, A. F. & Ulmer, R. G. 1995. Analysis of fatal motorcycle crashes: crash typing. *Accident Analysis & Prevention,* 27**,** 845-851.
Quddus, M. A., Noland, R. B. & Chin, H. C. 2002. An analysis of motorcycle injury and vehicle damage severity using ordered probit models. *Journal of Safety research,* 33**,** 445-462.
Rajalin, S. 1994. The connection between risky driving and involvement in fatal accidents. *Accident Analysis & Prevention,* 26**,** 555-562.
Rifaat, S. M., Tay, R. & De Barros, A. 2012. Severity of motorcycle crashes in Calgary. *Accident Analysis & Prevention,* 49**,** 44-49.

Rothman, L., Howard, A., Buliung, R., Macarthur, C., Richmond, S. A. & Macpherson, A. 2017. School environments and social risk factors for child pedestrian-motor vehicle collisions: a case-control study. *Accident Analysis & Prevention,* 98**,** 252-258.

Rowden, P., Watson, B., Haworth, N., Lennon, A., Shaw, L. & Blackman, R. 2016. Motorcycle riders' self-reported aggression when riding compared with car driving. *Transportation research part F: traffic psychology and behaviour,* 36**,** 92-103.

Savolainen, P. & Mannering, F. 2007. Probabilistic models of motorcyclists' injury severities in single- and multi-vehicle crashes. *Accident Analysis & Prevention,* 39**,** 955-963.

Schneider, W. H., Savolainen, P. T., Van Boxel, D. & Beverley, R. 2012. Examination of factors determining fault in two-vehicle motorcycle crashes. *Accident Analysis & Prevention,* 45**,** 669-676.

Shaheed, M. S. B., Gkritza, K., Zhang, W. & Hans, Z. 2013. A mixed logit analysis of two-vehicle crash severities involving a motorcycle. *Accident Analysis & Prevention,* 61**,** 119-128.

Shankar, V. & Mannering, F. 1996. An exploratory multinomial logit analysis of single-vehicle motorcycle accident severity. *Journal of safety research,* 27**,** 183-194.

Sosin, D. M., Sacks, J. J. & Holmgreen, P. 1990. Head injury—associated deaths from motorcycle crashes: relationship to helmet-use laws. *Jama,* 264**,** 2395-2399.

Stephens, A., Brown, J., De Rome, L., Baldock, M., Fernandes, R. & Fitzharris, M. 2017. The relationship between Motorcycle Rider Behaviour Questionnaire scores and crashes for riders in Australia. *Accident Analysis & Prevention,* 102**,** 202-212.

Tay, R. 2016. Comparison of the binary logistic and skewed logistic (Scobit) models of injury severity in motor vehicle collisions. *Accident Analysis & Prevention,* 88**,** 52-55.

Wali, B., Ahmed, A. & Ahmad, N. 2017. An ordered-probit analysis of enforcement of road speed limits. *Proceedings of the Institution of Civil Engineers-Transport***,** 1-10. http://dx.doi.org/10.1680/jtran.16.00141.

Wali, B., Khattak, A., David, G. & Liu, J. 2018a. Fuel Economy Gaps Within & Across Garages: A Bivariate Random Parameters Seemingly Unrelated Regression Approach (forthcoming). *International Journal of Sustainable Transportation*. 10.1080/15568318.2018.1466222.

Wali, B., Khattak, A., Waters, J., Chimba, D. & Li, X. 2018b. Development of Safety Performance Functions for Tennessee: Unobserved Heterogeneity & Functional Form Analysis (Accepted for publication). *Transportation Research Record: Journal of the Transportation Research Board* 10.1177/0361198118767409

Wali, B., Khattak, A. J., Bozdogan, H. & Kamrani, M. 2018c. How is Driving Volatility Related to Intersection Safety? A Bayesian Heterogeneity-Based Analysis of Instrumented Vehicles Data (Accepted for Publication). *Transportation Research Part C: Emerging Technologies*.

Wali, B., Khattak, A. J. & Karnowski, T. 2018d. How Driving Volatility in Time to Collision Relates to Crash Severity in a Naturalistic Driving Environment? *Presented at the Transportation Research Board 97th Annual Meeting, Washington DC, 2018.*

Wali, B., Khattak, A. J. & Xu, J. 2018e. Contributory fault and level of personal injury to drivers involved in head-on collisions: application of copula-based bivariate ordinal models. *Accident Analysis & Prevention,* 110**,** 101-114.

Washington, S. P., Karlaftis, M. G. & Mannering, F. 2010. *Statistical and econometric methods for transportation data analysis*, CRC press.

Wells, S., Mullin, B., Norton, R., Langley, J., Connor, J., Jackson, R. & Lay-Yee, R. 2004. Motorcycle rider conspicuity and crash related injury: case-control study. *Bmj,* 328**,** 857.

Zhao, S. & Khattak, A. 2015. Motor vehicle drivers' injuries in train–motor vehicle crashes. *Accident Analysis & Prevention,* 74**,** 162-168.

Zhao, S. & Khattak, A. J. 2017. Injury severity in crashes reported in proximity of rail crossings: The role of driver inattention. *Journal of Transportation Safety & Security***,** 1-18.